\documentclass{article}
\usepackage[latin9]{inputenc}
\usepackage{mathtools}
\usepackage{bm}
\usepackage{amsmath}
\usepackage{amssymb}
\usepackage{stmaryrd}
\usepackage{graphicx}

\makeatletter

\usepackage[level=0]{wgroup_message}  
\usepackage{algorithm}
\usepackage{algpseudocode}
\usepackage{bm}      

\usepackage{spconf}
\usepackage{color} 
\usepackage{graphicx,psfrag,cite}
\usepackage{amssymb}
\usepackage{amsmath, nccmath}
\usepackage{stmaryrd}
\usepackage[acronym]{glossaries}
\newcommand{\newac}{\newacronym}
\newcommand{\ac}{\gls}

\makeglossaries

\newac{speb}{SPEB}{square position error bound}
\newac[plural=EFIMs,firstplural=Fisher information matrices (EFIMs)]{efim}{EFIM}{Fisher information matrix}
\newac{ne}{NE}{Nash equilibrium}
\newac{mse}{MSE}{mean squared error}
\newac{toa}{TOA}{time of arrival}
\newac{tdoa}{TDOA}{time difference of arrival}
\newac{snr}{SNR}{signal-to-noise ratio}
\newac{lan}{LAN}{local area network}
\newac{psd}{PSD}{positive semidefinite}
\newac{pd}{PD}{positive definite}
\newac{wrt}{w.r.t.}{with respect to}
\newac{lhs}{L.H.S.}{left hand side}
\newac{wp1}{w.p.1}{with probability 1}
\newac{kkt}{KKT}{Karush-Kuhn-Tucker}
\newac{wlog}{w.l.o.g.}{without loss of generality}
\newac{mle}{MLE}{maximum likelihood estimation}
\newac{rssi}{RSSI}{received signal strength indicator}
\newac{mimo}{MIMO}{multiple-input multiple-output}
\newac{csi}{CSI}{channel state information}
\newac{fdd}{FDD}{frequency division duplexing}
\newac{ms}{MS}{mobile station}
\newac{bs}{BS}{base station}
\newac{bss}{BSs}{base stations}
\newac{d2d}{D2D}{device-to-device}
\newac{slnr}{SLNR}{signal-to-interference-leakage-and-noise-ratio}
\newac{ula}{ULA}{uniform linear antenna array}
\newac{pas}{PAS}{power angular spectrum}
\newac{mmse}{MMSE}{minimum mean square error}
\newac{zf}{ZF}{zero-forcing}
\newac{rzf}{RZF}{regularized zero-forcing}
\newac{as}{AS}{angular spread}
\newac{ue}{UE}{user equipment}
\newac{ues}{UEs}{user equipments}
\newac{iid}{i.i.d.}{independent and identically distributed} 
\newac{sinr}{SINR}{signal-to-interference-and-noise ratio}
\newac{tdd}{TDD}{time-division duplex}
\newac{rvq}{RVQ}{random vector quantization}
\newac{rhs}{R.H.S.}{right hand side}
\newac{mrc}{MRC}{maximum ratio combining}
\newac{cdf}{CDF}{cumulative distribution function}
\newac{a.s.}{a.s.}{almost surely}
\newac{los}{LOS}{line-of-sight}
\newac{jsdm}{JSDM}{joint spatial division and multiplexing}
\newac{map}{MAP}{maximum a posteriori}
\newac{klt}{KLT}{Karhunen-Lo\`eve Transform}
\newac{lbe}{LBE}{link bargaining equilibrium}
\newac{se}{SE}{Stackelberg equilibrium}
\newac{uav}{UAV}{unmanned aerial vehicle}
\newac{uavs}{UAVs}{unmanned aerial vehicles}
\newac{nlos}{NLOS}{non-line-of-sight}
\newac{pdf}{PDF}{probability density function}
\newac{em}{EM}{expectation-maximization}
\newac{knn}{KNN}{$k$-nearest neighbors}
\newac{svd}{SVD}{singular value decomposition}
\newac{nmf}{NMF}{non-negative matrix factorization}
\newac{umf}{UMF}{unimodality-constrained matrix factorization}
\newac{rmse}{RMSE}{rooted mean squared error}
\newac{olos}{OLOS}{obstructed line-of-sight}
\newac{mmw}{mmW}{millimeter wave}
\newac{ber}{BER}{bit error rate}
\newac{rss}{RSS}{received signal strength}
\newac{lp}{LP}{linear program}
\newac{ufw}{U-FW}{unimodal Frank-Wolfe}
\newac{utf}{UTF}{unimodality-constrained tensor factorization}
\newac{fw}{FW}{Frank-Wolfe}
\newac{iot}{IoT}{Internet-of-Things}
\newac{mae}{MAE}{mean absolute error}
\newac{crb}{CRB}{Cram\'er-Rao bound}
\newac{aoa}{AoA}{angle of arrival}
\newac{wcl}{WCL}{weighted centroid localization}
\newac{slf}{SLF}{spatial loss function}
\newac{btd}{BTD}{block-term tensor decomposition}
\newac{tps}{TPS}{thin plate spline}
\newac{nmse}{NMSE}{normalized mean squared error}
\newac{svt}{SVT}{singular value thresholding}
\newac{vae}{VAE}{variational autoencoder}
\newac{gan}{GAN}{generative adversarial network}
\newac{aod}{AOD}{angle of departure}
\newac{rbf}{RBF}{radial basis function}
\newac{loocv}{LOOCV}{leave one out cross validation}
\newac{lpr}{LPR}{local polynomial regression}
\newac{xl-mimo}{XL-MIMO}{extremely large multiple-input multiple-output}
\newac{6g}{6G}{sixth-generation}
\newac{cpd}{CPD}{canonical polyadic decomposition}
\newac{nnm}{NNM}{nuclear norm minimization}
\newac{mimo-ofdm}{MIMO-OFDM}{multiple-input multiple-output orthogonal frequency division multiplexing}
\newac{mimo-cdma}{MIMO-CDMA}{multiple-input multiple-output code division multiple access}
\newac{pad}{PAD}{power-angular-delay}
\newac{hmm}{HMM}{hidden Markov model}
\newac{dft}{DFT}{discrete time Fourier transform}
\newac{pnp}{PnP}{Plug-and-Play}
\newac{ode}{ODE}{ordinary differential equation}
\newac{gps}{GPs}{Gaussian processes}
\newac{ai}{AI}{artificial intelligence}
\newac{2d}{2D}{two-dimensional}
\usepackage{color}
\usepackage{graphicx}
\usepackage{psfrag}\usepackage{cite}

\ninept
\name{Hao~Sun\textsuperscript{1}, Shicong~Liu\textsuperscript{1}, Xianghao Yu\textsuperscript{1}, Ying Sun\textsuperscript{2}}
\vspace{-0.5em}
\address{\textsuperscript{1}City University of Hong Kong, Hong Kong, \textsuperscript{2}Pennsylvania State University, PA, USA}
\usepackage{bm}
\usepackage{bm}

\newac{uaps}{UAPS}{utility-aware path search}

\makeatother

\begin{document}
\title{Flow Matching-Based Active Learning for Radio Map Construction with
Low-Altitude UAVs}
\maketitle
\begin{abstract}
The employment of \ac{uavs} in the low-altitude economy necessitates
precise and real-time radio maps for reliable communication and safe
navigation. However, constructing such maps is hindered by the infeasibility
of exhaustive measurements due to \ac{uavs}' limited flight endurance.
To address this, we propose a novel active learning framework for
low-altitude radio map construction based on limited measurements.
First, a \ac{pnp}-refined flow matching algorithm is introduced,
which leverages flow matching as a powerful generative prior within
a \ac{pnp} scheme to reconstruct high-fidelity radio maps. Second,
the generative nature of flow matching is exploited to quantify uncertainty
by generating an ensemble of radio maps and computing the location-wise
variance. The resulting uncertainty map guides a multi-objective candidate
selection and then a trajectory is planned via \ac{uaps}, directing
the UAV to the most informative locations while taking travel costs
into account. Simulation results demonstrate that our method significantly
outperforms the baselines, achieving more than a 70\% reduction in
\ac{nmse}.

\begin{keywords}Active learning, flow matching, Plug-and-Play, radio
map, UAVs.\end{keywords}
\end{abstract}

\section{Introduction}

The rapid development of the low-altitude economy, empowered by the
widespread deployment of \ac{uavs} in logistics, surveillance, and
wireless communications \cite{WuMWuHLuW:J25,CheLiBSunCui:J25}, has
created a strong need for accurate radio maps \cite{TimShrFux:J24,XuLCheChePuw:C25}.
A radio map reveals the spatial distribution of channel characteristics
such as \ac{rss}, which plays a vital role in ensuring reliable communication,
efficient resource allocation, and safe UAV operations, especially
in complex urban environments \cite{MoxHuaXuj:J19,LiBChe:J24}.

However, building such radio maps is difficult. The large areas to
be covered, combined with the limited flight time and battery capacity
of UAVs, make exhaustive measurements infeasible. Thus, reconstructing
a complete radio map calls for advanced sampling and construction
methods.

Existing approaches can be categorized into two groups. Traditional
methods leverage statistical models such as \ac{gps} to interpolate
sparse data. While GPs can estimate prediction uncertainty and provide
guidance for sampling \cite{CheZhuWanLin:C25,PolSadYeW:J24}, they
fail to scale and adapt to complex scenarios. More recent studies
utilize deep learning techniques, including autoencoders and U-Nets,
to learn latent map structures from data \cite{CheWanGuo:J25,LuWGaoWenLia:C25,ShrRomChe:J23}.
Unlike GPs, these models do not naturally provide reliable uncertainty
estimates, which are essential for guiding sampling in active learning.
To address this, these models typically incorporate additional modules
for uncertainty estimation, such as Bayesian neural networks or deep
ensembles. However, these methods are computationally expensive and
may not yield reliable uncertainty estimates.

In recent years, the emergence of generative \ac{ai} offers an alternative,
as its inherent stochasticity allows for the diverse outputs that
naturally contain uncertainty, which serves as a direct and efficient
measure of uncertainty. Yet, a unified framework that leverages generative
diversity for both high-quality radio map reconstruction and uncertainty
quantification remains to be explored.
\begin{figure}
\includegraphics[width=1\columnwidth]{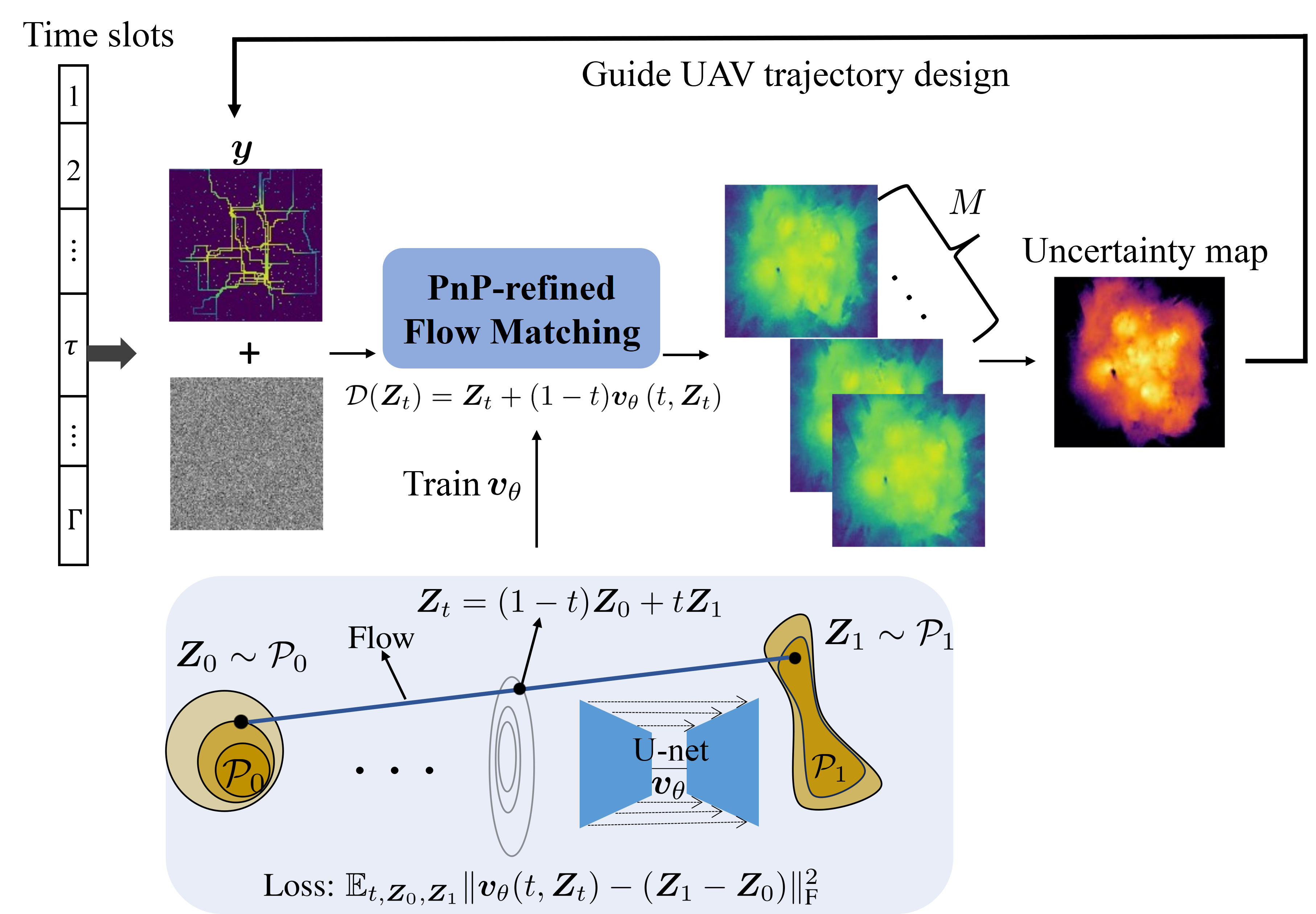}\caption{\label{fig:Active-learning-process}Flow matching-based active learning
process for radio map construction.}
\end{figure}

In this paper, we develop a flow matching-based active learning framework
for UAV-based radio map construction (cf. Fig.~\ref{fig:Active-learning-process}).
We embed flow matching as a generative prior within a \ac{pnp} scheme
\cite{ZhaLiYZuo:J22}, and then introduce an adaptive refinement step
that polishes the reconstruction in the final stages for higher fidelity.
To obtain the uncertainty map, we exploit the generative nature of
flow matching to generate multiple plausible radio maps and compute
their location-wise variance. This uncertainty is then exploited to
guide a probabilistic sampling strategy, which directs the UAV to
informative yet cost-efficient measurement locations.

Our main contributions are summarized as follows:
\begin{itemize}
\item We propose a PnP-refined flow matching reconstruction scheme whose
denoiser is defined by the flow matching velocity field and augmented
with an adaptive inner-loop refinement.
\item We quantify the uncertainty via a generative radio map ensembles obtained
by repeatedly running the reconstruction process and computing the
variance at each location.
\item We design a probabilistic multi-objective selection rule that balances
informativeness and reachability, and plan a \ac{uaps} whose step
cost embeds the normalized uncertainty.
\end{itemize}

\section{Problem Statement}

We consider a low-altitude urban environment where multiple \ac{bss}
are deployed, either at ground level or on building rooftops. The
objective is to estimate the spatial distribution of the \ac{rss}
over a low-altitude \ac{2d} horizontal area, denoted by $\mathcal{S}\subset\mathbb{R}^{2}$,
using a UAV. This area is discretized into a uniform grid of size
$I\times J$, and the RSS value at each grid coordinate $(i,j)$ is
represented by the entry $\mathbf{M}_{ij}$ of a complete but unknown
radio map matrix $\mathbf{M}\in\mathbb{R}^{I\times J}$.

A UAV is deployed to reconstruct $\mathbf{M}$ based on a limited
number of measurements $\bm{y}$ collected along a trajectory across
discrete time slots indexed by $\tau=1,\dots,\Gamma$. In each time
slot $\tau$, an ensemble of $M$ candidate radio maps $\{\bm{Z}^{(\tau,m)}\}_{m=1}^{M}$
is generated by using a trained flow matching model based on all previously
acquired measurements $\bm{y}^{(\tau)}$. From this ensemble of radio
maps, an uncertainty map is derived to quantify the reliability of
the estimated RSS values at each grid point. The UAV exploits this
uncertainty map to determine its measurement trajectory in the next
time slot $\tau+1$.

\section{PnP-Refined Flow Matching for Radio Map Construction}

In this section, we study how to generate a candidate radio map $\bm{Z}$
in each time slot $\tau$. For notation simplicity, the dependence
on $\tau$ will be omitted throughout this section.

\subsection{Radio Map Construction via a PnP Framework}

The goal is to recover an unknown radio map $\bm{Z}$ from sparse
RSS measurements $\bm{y}\in\mathbb{R}^{Q}$. This can be formulated
as a \ac{map} estimation
\begin{equation}
\arg\max_{\bm{Z}}\left\{ \log f(\bm{y}|\bm{Z})+\log p(\bm{Z})\right\} ,\label{eq:map_formulation}
\end{equation}
where $f(\bm{y}|\bm{Z})$ is the likelihood, and $p(\bm{Z})$ is the
prior probability density function of the target radio map $\bm{Z}$.

Since the closed form of $p(\bm{Z})$ is unknown, we adopt the \ac{pnp}
framework, which approximates the prior $\log p(\bm{Z})$ with a learned
denoiser $\mathcal{D}(\cdot)$ . The reconstruction is then obtained
through the PnP gradient-denoising iteration \cite{ZhaLiYZuo:J22}
\begin{equation}
\begin{cases}
\hat{\bm{Z}}^{(u)}=\bm{Z}^{(u-1)}-\gamma\nabla F(\bm{Z}^{(u-1)};\bm{y})\\
\bm{Z}^{(u)}=\mathcal{D}(\hat{\bm{Z}}^{(u)})
\end{cases},\label{eq:pnp_iteration_detailed}
\end{equation}
where the data-fidelity term $F(\bm{Z};\bm{y})\coloneqq-\log f(\bm{y}|\bm{Z})$.
One example of $F$ is $F(\bm{Z};\bm{y})=\|\mathcal{H}(\bm{Z})-\bm{y}\|^{2}$,
with $\mathcal{H}:\mathbb{R}^{I\times J}\to\mathbb{R}^{Q}$ denoting
a degradation operator. The symbol $\nabla$ denotes the gradient
operator, and $\gamma$ is the step size. The first step enforces
measurement consistency, and the second step projects the estimation
onto the manifold of ground truth radio maps.

The performance of PnP crucially depends on the denoiser $\mathcal{D}(\cdot)$,
for which we employ flow matching as a powerful generative prior.

\subsection{Flow Matching}

Let $\mathcal{\mathcal{P}}_{1}$ represent the target distribution
over the space of radio maps $\mathbb{R}^{I\times J}$. Note that
in (\ref{eq:map_formulation}), $p(\bm{Z})$ refers to the density
function associated with $\mathcal{P}_{1}$. Assume that we only have
access to a number of radio map samples from $\mathcal{\mathcal{P}}_{1}$,
but cannot access the distribution itself. The goal of flow matching
is to build a neural network that can generate new samples from $\mathcal{P}_{1}$,
given a training dataset of radio map samples. To address this, flow
matching builds a probability flow $\{\mathcal{P}_{t}\}_{0\leq t\leq1}$,
from a known source distribution $\mathcal{P}_{0}$, typically a Gaussian
one \cite{LipCheBenNic:J23}, to the target distribution of radio
map $\mathcal{P}_{1}$.

A straightforward choice of flow is the linear interpolation between
a source sample $\bm{Z}_{0}\sim\mathcal{P}_{0}$ and a target sample
$\bm{Z}_{1}\sim\mathcal{P}_{1}$
\begin{equation}
\bm{Z}_{t}=(1-t)\bm{Z}_{0}+t\bm{Z}_{1},\quad t\in[0,1].\label{eq:-2}
\end{equation}
The instantaneous velocity of this path is $u(t,\bm{Z}_{t})=\mathrm{d}(\bm{Z}_{t})/\mathrm{d}t=\bm{Z}_{1}-\bm{Z}_{0}$.
Flow matching trains a neural network $\bm{v}_{\theta}(t,\bm{Z}_{t})$,
parameterized by $\theta$, to approximate $u(t,\bm{Z}_{t})$, with
the objective
\begin{equation}
\mathcal{L}(\theta)=\mathbb{E}_{t,\bm{Z}_{0},\bm{Z}_{1}}\|\bm{v}_{\theta}(t,\bm{Z}_{t})-(\bm{Z}_{1}-\bm{Z}_{0})\|_{\text{F}}^{2},\label{eq:-1}
\end{equation}
where $t$ is uniformly distributed over $[0,1]$, and $\|\cdot\|_{\text{F}}$
denotes Frobenius norm. 

The learned velocity field $\bm{v}_{\theta}(t,\bm{Z}_{t})$ determines
a time-dependent flow $\bm{\phi}:[0,1]\times\mathbb{R}^{I\times J}\to\mathbb{R}^{I\times J}$
\begin{equation}
\frac{\mathrm{d}\bm{\phi}(t,\bm{Z}_{0})}{\mathrm{d}t}=\bm{v}_{\theta}(t,\bm{\phi}(t,\bm{Z}_{0})),\label{eq:ode}
\end{equation}
with initial condition $\bm{\phi}(0,\bm{Z}_{0})=\bm{Z}_{0}$. By integrating
(\ref{eq:ode}) from $t=0$ to $t=1$, the initial noise map $\bm{Z}_{0}$
is smoothly transported along the learned flow to a final state $\bm{\phi}(1,\bm{Z}_{0})=\bm{Z}_{1}$
that lies on the manifold of ground truth radio maps.

\subsection{PnP-Refined Flow Matching Framework}

To integrate the learned flow matching prior into the PnP framework
in (\ref{eq:pnp_iteration_detailed}), we propose the PnP-refined
flow matching framework detailed in \textbf{Algorithm \ref{alg:pnp_refined_flow}}.
The framework proceeds over $K$ outer iterations. To align these
discrete iteration steps with the continuous time variable of the
flow, we define a discrete time sequence $t_{k}=k/K\in[0,1]$.

After the velocity field $\bm{v}_{\theta}(t,\bm{Z}_{t})$ is learned
in (\ref{eq:-1}), we define the denoiser $\mathcal{D}(\cdot)$ at
each time step $t_{k}$ as \cite{MarGagHag:J25}
\begin{equation}
\mathcal{D}(\tilde{\bm{Z}}_{t_{k}}^{(u)})\triangleq\tilde{\bm{Z}}_{t_{k}}^{(u)}+(1-t_{k})\bm{v}_{\theta}\left(t_{k},\tilde{\bm{Z}}_{t_{k}}^{(u)}\right).\label{eq:denoiser_def}
\end{equation}

 Each outer iteration $k$, corresponding to the time variable $t_{k}$,
contains an inner refinement loop of $U_{k}$ steps. Crucially, the
number of refinements $U_{k}$ increases with $k$, so that early
iterations with fewer inner steps rapidly establish the coarse structure
of the radio map, while later iterations with more inner steps gradually
refine the estimate to recover fine details. This progressive refinement
schedule is implemented through the following inner loop, which balances
data fidelity and the learned generative prior.

Each of the $U_{k}$ steps begins by steering the current estimate
$\bm{Z}_{t_{k}}^{(u-1)}$ toward consistency with the measurements
$\bm{y}$ via a gradient descent step on the data-fidelity term $F$
as in (\ref{eq:pnp_iteration_detailed}). To mitigate artifacts and
keep the trajectory aligned with the generative path, the intermediate
result $\hat{\bm{Z}}_{t_{k}}^{(u)}$ from the data-consistency update
is then regularized through Path Projection by interpolating it with
the initial noise map $\bm{Z}_{0}$: $\tilde{\bm{Z}}_{t_{k}}^{(u)}\leftarrow t_{k}\hat{\bm{Z}}_{t_{k}}^{(u)}+(1-t_{k})\bm{Z}_{0}$.
Finally, the Denoising step applies the learned denoiser $\mathcal{D}$
in (\ref{eq:denoiser_def}), which leverages the model's velocity
field to project the estimate onto the manifold of ground truth radio
maps, effectively removing structural inconsistencies introduced by
the data-consistency update.\begin{algorithm}[t]
\caption{PnP-Refined Flow Matching for Radio Map Reconstruction}
\label{alg:pnp_refined_flow}
\begin{algorithmic}[1]
\State \textbf{Input:} Observed RSS $\bm{y}$; an initial noise map $\bm{Z}_0\sim \mathcal{P}_0$; number of outer steps $K$; step size $\gamma$; refinement schedule $\{U_k\}_{k=1}^K$
\State \textbf{Output:} Reconstructed map $\bm{Z}$
\For{$k=1,\dots,K$} \Comment{outer loop at $t_k = k/K$}
    \State $t_k \gets k/K$
    \State $\bm{Z}_{t_k}^{(0)} \gets \bm{Z}_{t_{k-1}}$ 
    \For{$u=1,2,\dots,U_k$} \Comment{adaptive inner refinement loop}
        \State $\hat{\bm{Z}}_{t_k}^{(u)} \gets \bm{Z}_{t_k}^{(u-1)} - \gamma\,\nabla F\!\left(\bm{Z}_{t_k}^{(u-1)};\bm{y}\right)$
\Statex \Comment{data-consistency update}
        \State $\tilde{\bm{Z}}_{t_k}^{(u)} \gets t_k\,\hat{\bm{Z}}_{t_k}^{(u)} + (1-t_k)\,\bm{Z}_0$ \Comment{path projection}
        \State $\bm{Z}_{t_k}^{(u)} \gets \mathcal{D}_{t_k}\!\big(\tilde{\bm{Z}}_{t_k}^{(u)}\big)$ \Comment{apply the learned denoiser}
    \EndFor
    \State $\bm{Z}_{t_k} \gets \bm{Z}_{t_k}^{(U_k)}$ \Comment{result of outer step $k$}
\EndFor
\State \Return $\bm{Z} \gets \bm{Z}_{t_K}$
\end{algorithmic}
\end{algorithm}

\section{Active Learning via Uncertainty Quantification}

With a powerful PnP-refined flow matching framework for reconstruction,
the next critical task is to intelligently guide the UAV's trajectory
for the sampling in the next time slot $\tau+1$.

\subsection{Uncertainty Quantification with Generative Ensembles}

A key feature of the PnP-refined framework is its generative capability,
which enables the construction of an ensemble of plausible radio maps
rather than a single reconstruction, all consistent with the measurements
$\bm{y}^{(\tau)}$ collected up to time slot $\tau$. Specifically,
an ensemble of $M$ reconstructions $\{\bm{Z}^{(\tau,m)}\}_{m=1}^{M}$
is generated by executing Algorithm 1 for $M$ times, each initialized
with a sample $\bm{Z}_{0}\sim\mathcal{P}_{0}$.

Although each map is consistent with the acquired measurements, discrepancies
arise in unmeasured regions where the model depends on its learned
prior. The variation across ensemble members thus provides a principled
measure of model uncertainty. The uncertainty map in time slot $\tau$
is then defined as the location-wise variance across the ensemble
as
\begin{equation}
\bm{U}^{(\tau)}=\frac{1}{M}\sum_{m=1}^{M}\left(\bm{Z}^{(\tau,m)}-\frac{1}{M}\sum_{n=1}^{M}\bm{Z}^{(\tau,n)}\right)^{2}.\label{eq:-3}
\end{equation}

\subsection{Active Sampling Strategy}

Based on the uncertainty map $\bm{U}^{(\tau)}$, we first select several
$N$ candidate informative locations to be sampled. Then, we propose
a \ac{uaps} approach to direct the UAV flying trajectory passing
through all $N$ candidates in the next time slot.

\subsubsection{Candidate Locations Selection}

We define $\mathcal{V}^{(\tau)}$ as the set of sampled locations,
and $\mathcal{V}_{\text{unsampled}}^{(\tau)}$, as the set of unsampled
locations in time slot $\tau$.

A straightforward approach is to always select the location with the
maximum uncertainty \cite{ShrRomChe:J23,PolSadYeW:J24}. However,
such a greedy rule is prone to myopic behavior, as the UAVs may repeatedly
fly to isolated high-variance points, thereby lengthening the flight
trajectory, increasing energy consumption, and ultimately leading
to inefficient trajectories.

To address this, we adopt a probabilistic multi-objective strategy.
Instead of committing to a single maximizer, we sample a diverse set
of candidate objectives that balance informativeness with reachability.
Concretely, a set of $N$ candidate locations $\mathcal{G}^{(\tau)}=\{\bm{g}_{1}^{(\tau)},\ldots,\bm{g}_{N}^{(\tau)}\}$
is drawn without replacement from $\mathcal{V}_{\text{unsampled}}^{(\tau)}$,
where $\mathbf{g}_{n}^{(\tau)}=(i_{n}^{(\tau)},j_{n}^{(\tau)})$ is
the grid coordinate. The selection criterion is given by the weight
\begin{equation}
w(\bm{g}_{n}^{(\tau)})=\frac{\bm{U}_{i_{n}j_{n}}^{(\tau)}}{1+\kappa d\left(\bm{g}_{\text{UAV}}^{(\tau)},\bm{g}_{n}^{(\tau)}\right)},\label{eq:probability}
\end{equation}
where $\bm{U}_{i_{n}j_{n}}^{(\tau)}$ is the uncertainty at location
$\bm{g}_{n}^{(\tau)}$, $d(\cdot,\cdot)$ is the Manhattan distance,
$\kappa\ge0$ trades off informativeness $\bm{U}_{i_{n}j_{n}}^{(\tau)}$
versus $d(\cdot,\cdot)$, and $\bm{g}_{\text{UAV}}^{(\tau)}$ denotes
the initial UAV position in time slot $\tau$.

\subsubsection{Visiting Order and Overall Trajectory Planning}

In the following, we omit the superscript $(\tau)$ for notational
simplicity. Given the $N$ candidate locations in $\mathcal{G}$,
we need to determine an efficient visiting order when designing trajectory.
Let $\Pi(\mathcal{G})$ denote the set of all permutations of $\mathcal{G}$.
For a given visiting order $\sigma\in\Pi(\mathcal{G})$, the total
trajectory cost is
\begin{equation}
J(\sigma)=C\!\left(\bm{g}_{\text{UAV}},\bm{g}_{\sigma(1)}\right)+\sum_{i=1}^{N-1}C\!\left(\bm{g}_{\sigma(i)},\bm{g}_{\sigma(i+1)}\right).\label{eq:}
\end{equation}
The trajectory cost between two candidate locations is given by
\begin{equation}
C(\bm{g}_{0},\bm{g}_{L})=\min_{\pi:\,\bm{g}_{0}\to\bm{g}_{L}}\;\sum_{l=0}^{L-1}c(\bm{g}_{l},\bm{g}_{l+1}),\label{eq:path_optimization}
\end{equation}
where $\bm{g}_{0}$ denotes the start candidate location, i.e., $\bm{g}_{\text{UAV}}$
or $\bm{g}_{\sigma(i)}$ in (\ref{eq:}) and $\bm{g}_{L}$ denotes
the end candidate location, i.e., $\bm{g}_{\sigma(1)}$ or $\bm{g}_{\sigma(i+1)}$
in (\ref{eq:}), and $\pi:\bm{g}_{0}\to\bm{g}_{L}$ denotes an arbitrary
trajectory from $\bm{g}_{0}$ to $\bm{g}_{L}$. The per-step cost
$c\bigl(\bm{g}_{l},\bm{g}_{l+1}\bigr)$ is designed to reward the
exploration of uncertain regions
\begin{equation}
c\bigl(\bm{g}_{l},\bm{g}_{l+1}\bigr)=1-\beta\frac{\bm{U}_{i_{l+1}j_{l+1}}-\text{min}(\bm{U})}{\text{max}(\bm{U})-\text{min}(\bm{U})},\label{eq:step_cost}
\end{equation}
where $\beta\in[0,1]$ is a parameter controlling the incentive for
exploration, and $\text{min}(\bm{U})$ and $\text{max}(\bm{U})$ denote
the minimum and maximum values among all entries of $\bm{U}$, respectively.
A higher uncertainty $\bm{U}_{i_{l+1}j_{l+1}}$ at the next step $\bm{g}_{l+1}$
leads to a lower step cost $c\bigl(\bm{g}_{l},\bm{g}_{l+1}\bigr)$,
encouraging the trajectory to pass through such informative locations.

The optimal order and trajectory can be obtained by $\sigma^{*}=\arg\underset{\sigma\in\Pi(\mathcal{G})}{\min}J(\sigma).$

\subsubsection{UAPS for Trajectory Planning Between Two Candidates}

Directly solving for \eqref{eq:path_optimization} by enumerating
all possible trajectories is computationally intractable. Therefore,
we propose an efficient search algorithm named \ac{uaps} to find
the optimal trajectory $\pi:\bm{g}_{0}\to\bm{g}_{L}$ from a starting
point $\bm{g}_{0}$ to an objective $\bm{g}_{L}$.

Starting from the UAV position $\bm{g}_{0}$, UAPS repeatedly examines
the neighboring grids. For each neighbor location $\bm{g}$, the evaluation
function is defined as $\varphi(\bm{g})=r(\bm{g})+h(\bm{g}).$ Here,
$r(\bm{g})$ is the actual cost of the trajectory from the start location
$\bm{g}_{0}$ to the candidate location $\bm{g}$, calculated by summing
the per-step costs $c\bigl(\bm{g}_{l},\bm{g}_{l+1}\bigr)$ in (\ref{eq:step_cost})
along the trajectory. The term $h(\bm{g})$ is a heuristic estimate
of the remaining flying cost from the candidate location $\bm{g}$
to the objective $\bm{g}_{L}$, calculated as $h(\bm{g})=(1-\beta)d(\bm{g},\bm{g}_{L})$.
This formulation ensures that our search for an optimal trajectory
correctly balances trajectory length and information gain in a computationally
feasible manner.

At each location planning, UAPS expands the location with the smallest
$\varphi(\bm{g})$, explores its feasible neighbors, and updates their
costs. This process continues until the objective location $\bm{g}_{L}$
is reached. The resulting trajectory yields the travel cost (\ref{eq:path_optimization}),
which is then employed in (\ref{eq:}) to determine the optimal visiting
order $\sigma^{*}$.
\begin{figure}
\includegraphics[width=1\linewidth]{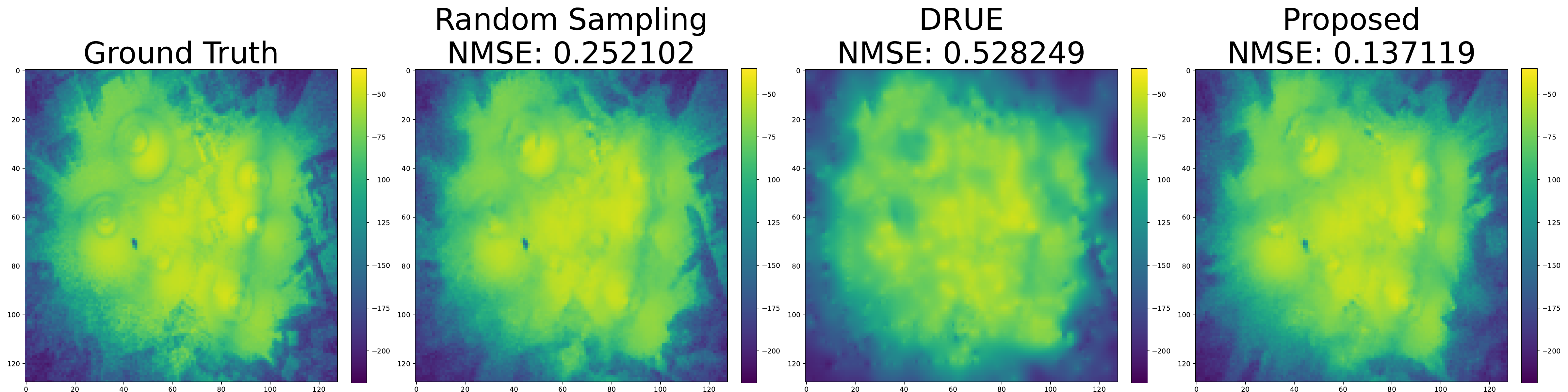}\caption{\label{fig:Visual-comparison-of}Visual comparison of reconstruction
radio maps.}
\end{figure}

\section{Simulation Results}

In this section, we present the simulation results to validate the
effectiveness of our proposed flow matching-based active learning
framework.
\begin{figure}
\includegraphics[width=1\columnwidth]{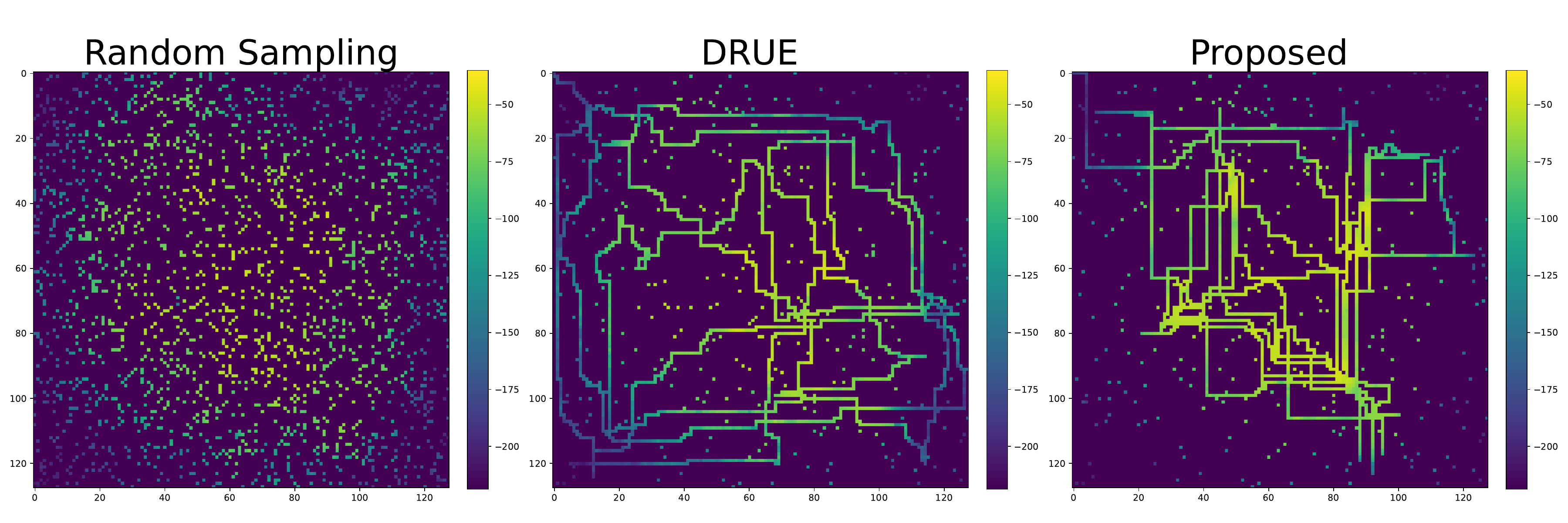}

\caption{\label{fig:trajectory}Sampling patterns and UAV trajectories for
Random Sampling, DRUE, and the proposed method.}
\end{figure}

The ground truth radio maps were generated using the Sionna-RT ray-tracing
simulator \cite{FayJakMer:J25}, which provides physically accurate
radio channel predictions. We adopt the open-source ``Etoile'' 3D
urban scenario, and randomly deploy $7$ transmitters within this
scenario. The ground truth radio map is then computed over a planar
grid of $128\times128$ at a fixed altitude of $50$ m, yielding an
RSS matrix $\mathbf{M}$ with dimensions $I=J=128$. This process
is repeated to generate a dataset of $800$ radio maps for training
and $30$ radio maps for testing. In the training stage of flow matching,
the source distribution $\mathcal{P}_{0}$ is chosen as $\mathcal{N}(0,1)$.

For demonstration, we assume that $2\%$ of the grid locations are
initially observed. The UAV then actively samples a total of $2,000$
additional locations, guided by the proposed active learning strategy.
The degradation operator $\mathcal{H}$ models the UAV measurement
process. $\mathcal{H}(\bm{Z})$ extracts the entries of the radio
map corresponding to the locations that the UAV has visited. The PnP-refined
flow matching algorithm is configured with $K=50$ iterative steps
and step size $\gamma=2$. For steps $k<46$, the refinement step
$U_{k}$ is set to $1$, and increases to $10$ in later iterations
to enhance reconstruction precision. In the active learning phase,
we set $\kappa=0.001$ in (\ref{eq:probability}) and exploration
incentive $\beta=0.9$ in (\ref{eq:step_cost}). The number of candidate
objectives per iteration is set to $N=10$, and the ensemble size
for uncertainty estimation $M=5$.

The performance of the reconstruction is quantified using the \ac{nmse}.
To ensure physical consistency, the NMSE is calculated in the linear
power domain rather than the dB scale. The NMSE is then computed as
$\text{NMSE}=\frac{\|10^{\mathbf{M}/10}-10^{\bm{Z}/10}\|_{\text{F}}^{2}}{\|10^{\mathbf{M}/10}\|_{\text{F}}^{2}},$
where $\mathbf{M}$ and $\bm{Z}$ are the ground truth and reconstructed
radio map in the linear power scale, respectively.

We benchmark our method against two sampling strategies, both constrained
to the same overall sampling locations. Baseline 1: Random Sampling.
In this baseline, $2,000$ additional sampling locations are selected
uniformly at random from the entire grid. The reconstruction is then
performed through the proposed PnP-refined flow matching based on
all collected samples. This strategy does not involve an iterative,
uncertainty-guided trajectory planning process. Baseline 2: Deep Radio
Map and Uncertainty Estimator (DRUE) \cite{ShrRomChe:J23}. This method
employs two autoencoders to estimate the radio map and its uncertainty
separately, followed by a UAV trajectory designed in an uncertainty-aware
manner.
\begin{figure}
\centering\includegraphics[width=0.78\columnwidth]{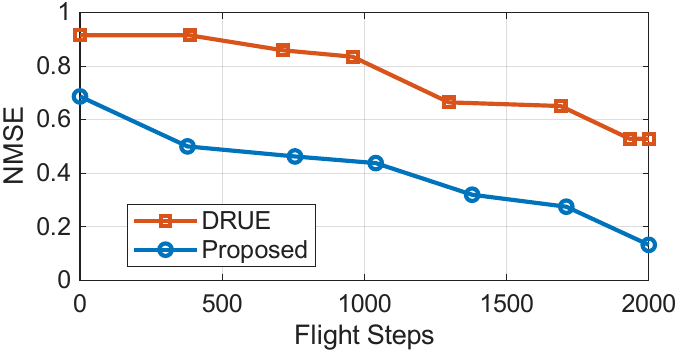}

\caption{\label{fig:Reconstruction-NMSE-versus sampling process}Reconstruction
NMSE versus the flight steps.}
\end{figure}

The evaluation is conducted from three perspectives: qualitative reconstruction
quality, sampling trajectory efficiency, and quantitative reconstruction
error.

Fig.~\ref{fig:Visual-comparison-of} provides a qualitative comparison
of the radio map reconstruction after $2,000$ active samples. A visual
inspection reveals that the radio map reconstructed by the proposed
method more closely resembles the ground truth in both its detailed
textures and overall spatial structure, qualitatively demonstrating
its superior performance.

The superior reconstruction quality is a direct consequence of our
intelligent sampling strategy, as illustrated in Fig.~\ref{fig:trajectory}.
While Random Sampling results in a scattered and inefficient coverage,
and DRUE broadly explores the entire area without concentrating on
truly informative regions, our method guides the UAV to precisely
target the most informative regions.

The quantitative improvement is demonstrated in Fig.~\ref{fig:Reconstruction-NMSE-versus sampling process}
which presents the reconstruction NMSE along the active learning process.
It is clear that the NMSE of the proposed method consistently decreases
at a faster rate than that of the DRUE baseline. Ultimately, our approach
achieves an NMSE improvement of over $70\%$ compared to the DRUE
baseline, representing a significant performance gain.

\section{Conclusion}

This paper introduced an active learning framework for UAV-based radio
map construction. A PnP-refined flow matching algorithm was introduced
to enable high-fidelity map recovery. Radio map uncertainty was quantified
via location-wise variance over generative ensembles, which subsequently
guides a utility-aware trajectory planner to optimize UAV sampling
trajectories. Simulation results verified that the proposed framework
achieves superior reconstruction accuracy compared with baseline methods.

\vfill{}

\bibliographystyle{IEEEtran}
\bibliography{IEEEabrv,StringDefinitions,JCgroup,ChenBibCV}

\end{document}